\documentclass[10pt,a4paper]{article}

\usepackage[margin=0.69in]{geometry}
\usepackage{authblk}
\usepackage{cite}
\usepackage{url}
\usepackage{tipa}
\usepackage[dvips]{graphicx}
\usepackage{amssymb}
\usepackage{color}
\usepackage{amsbsy}
\usepackage{textcomp}  
\usepackage{url}
\usepackage{booktabs}
\usepackage{multirow}
\usepackage{marvosym}

\usepackage[cmex10]{amsmath}
\usepackage{graphicx}
\usepackage{amssymb}
\usepackage{mathrsfs}
\usepackage{cleveref}
\usepackage{cite}

\newcommand{\bs}[1]{\boldsymbol{#1}}

\newcommand{\m}[1]{\mathcal{#1}}

\newcommand{\eop}{\hfill $\Box$ \\}
\newcommand{\eor}{\hfill $\bigtriangledown$ \\}

\newcommand{\proof}{\noindent \textit{Proof}: }

\newtheorem{lemma}{Lemma}
\newtheorem{remark}{Remark}

	{\begin{list}{}%
	{\setlength{\leftmargin}{#1}}%
	\item[]%
    			}
	{\end{list}}

\newcommand{\qed}{\nobreak \ifvmode \relax \else
      \ifdim\lastskip<1.5em \hskip-\lastskip
      \hskip1.5em plus0em minus0.5em \fi \nobreak
      \vrule height0.75em width0.5em depth0.25em\fi}
      
\usepackage{lipsum}

\newcommand\blfootnote[1]{%
  \begingroup
  \renewcommand\thefootnote{}\footnote{#1}%
  \addtocounter{footnote}{-1}%
  \endgroup
}

\title{\LARGE \textbf{A MAP approach for \MakeLowercase{$\ell_q-$}norm regularized sparse parameter estimation using the EM algorithm}$^\star$}

\author[$^\dagger$]{\Large Rodrigo Carvajal}
\author[$^\dagger$]{\Large Juan C. Ag\"uero}
\author[$^\ddagger$]{\Large Boris I. Godoy}
\author[$^\ast$]{\Large Dimitrios Katselis}
\affil[$^\dagger$]{\normalsize Electronics Engineering Department, Universidad T\'{e}cnica Federico Santa Mar\'ia, Chile. \Letter~\{rodrigo.carvajalg,~juan.aguero\}@usm.cl}
\affil[$^\ddagger$]{\normalsize School of Electrical Engineering and Telecommunications, The University of New South Wales, Australia. \Letter~{b.godoy@unsw.edu.au}}
\affil[$^\ast$]{\normalsize Coordinated Science Laboratory, University of Illinois at Urbana-Champaign, IL, USA. \Letter~{katselis@illinois.edu}}
\date{}

\begin{document}
\maketitle
\blfootnote{\textbf{$^\star$Accepted to IEEE Machine Learning for Signal Processing Conference 2015.}}
\blfootnote{The work of R. Carvajal was supported by Chile's National Fund for Scientific and Technological Development (FONDECYT) through its Post-doctoral Fellowship Programme 2014 -- Grant No 3140054.} \blfootnote{The work of J. C. Ag\"uero was supported by FONDECYT through grant No 1150954.}\blfootnote{The work of D. Katselis was partially supported by DTRA Grant HDTRA1-13-1-0030.}\blfootnote{This work was partially supported by the Scientific Technological Centre of Valpara\'iso (CCTVal, Proyecto Basal FB0821), Chile.}\blfootnote{\textbf{Corresponding author}: Rodrigo Carvajal.}
\hrule \hrule
\begin{abstract}
In this paper, Bayesian parameter estimation through the consideration of the Maximum A Posteriori (MAP) criterion is revisited under the prism of the Expectation-Maximization (EM) algorithm. By incorporating a sparsity-promoting penalty term in the cost function of the estimation problem through the use of an appropriate prior distribution,
we show how the EM algorithm can be used to efficiently solve the corresponding optimization problem. To this end, we rely on variance-mean Gaussian mixtures (VMGM) to describe the prior distribution, while we incorporate many nice features of these mixtures to our estimation problem. The corresponding MAP estimation problem is completely expressed in terms of the EM algorithm, which allows for handling nonlinearities and \textit{hidden variables} that cannot be easily handled with traditional methods. For comparison purposes, we also develop a Coordinate Descent algorithm for the $\ell_q$-norm penalized problem and present the performance results via simulations.
\end{abstract}\vspace{2mm}
\hrule \hrule
\section{Introduction}
\label{sec:intro}
In engineering, an overwhelming number of applications deal with relating a random response variable $\bs{y}$ with a set of explanatory variables or covariates $\bs{X}=\left(\bs{x}_1,\bs{x}_2,\ldots,\bs{x}_n\right)$ through a regression model. The most usual case is that of the linear regression, namely\vspace{-3mm}
\begin{equation}\label{eq:lGm}
\bs{y}=\bs{X}\bs{\theta}+\bs{\epsilon},\vspace{-3mm}
\end{equation}
where $\bs{y}$ is the $n\times 1$ response vector, $\bs{X}$ is the $n\times p$ regression matrix, $\bs{\theta}$ is the $p\times 1$ parameter vector and $\bs{\epsilon}$ is the $n\times 1$ noise vector. The traditional way to perform parameter estimation in this setup is by considering the Maximum Likelihood (ML) approach. The corresponding estimates can be iteratively computed based on the Expectation-Maximization (EM) algorithm  \cite{ref:Dempster1977}, when the observations are seen as \emph{incomplete} data, see, e.g., \cite{ref:McLachlan}. 

On the other hand, when the parameter vector $\bs \theta$ is sparse, i.e., many of its entries are equal to zero, it is common to include a sparsity promoting penalty to the ML problem, yielding a regularized ML estimation problem. Those sparsity promoting penalties are usually the $\ell_1$-norm, the $\ell_q$-norm ($0<q<1$) and the $\ell_0$ pseudo-norm \cite{ref:Donoho2006a, ref:Donoho2006b}. Depending on the penalty, different techniques are utilized to solve this regularized ML problem, such as \textit{proximal methods} (PM), see, e.g., \cite{ref:Baldassarre}, or the \textit{local linear approximation} (LLA) and the \textit{local quadratic approximation} (LQA) methods \cite{ref:Zou2008, ref:Fan2010}. Another optimization approach for penalized problems corresponds to the implementation of the Coordinate Descent (CD) algorithm \cite{ref:Zangwill}. This technique has been successfully applied to the Lasso problem \cite{ref:Wu2008}. Its extension to non-convex penalties was introduced in \cite{ref:Breheny2011}. Its particular application to sparsity problems with the $\ell_q$-norm penalty and its convergence is analysed in \cite{ref:Marjanovic2014}. In CD, the vector parameter estimates are obtained in a \textit{per coordinate} base. Within the EM framework, the application of the CD approach yields the so-called Expectation-Conditional Maximum (ECM) algorithm \cite{ref:McLachlan}. In this paper, we show how to extend the ECM algorithm to $\ell_q$-norm penalized ML problems with \textit{hidden variables} using the CD method.

A different approach corresponds to turn the penalty function in the form of a probability density function (pdf), accounting for some prior knowledge of the vector parameter $\bs \theta$. Thus, we can turn the ML estimation problem into a Maximum a Posteriori (MAP) problem. If the observations correspond to \textit{incomplete data}, the EM algorithm can be easily modified to produce the posterior mode of $\bs{\theta}$ instead of the ML estimate of $\bs{\theta}$. Denoting by $p(\bs{y}|\bs{\theta})$ the response sampling density, the corresponding log-likelihood function is $\mathcal{L}(\bs{\theta})=\log p(\bs{y}|\bs{\theta})$. For ML estimation, the EM algorithm iteratively selects a value $\bs{\theta}^{*}$ that maximizes $\mathcal{L}(\bs{\theta})$. In the case of MAP estimation, the function that is maximized by the EM algorithm is $\mathcal{L}(\bs{\theta})+G(\bs{\theta})$, where $G(\bs{\theta})=\log p(\bs{\theta})$ is the log-prior of the parameter vector $\bs{\theta}$. From these observations, it becomes clear that the EM algorithm is a quite general tool for solving either classical or Bayesian estimation problems.

In this paper, we formalize an EM-based framework for regularized ML parameter estimation using a MAP perspective. Motivated by the recent advances in regularized parameter estimation and in compressed sensing, we propose a way to introduce the $\ell_q-$norm sparsity-promoting penalty into our estimation problem by properly selecting the prior through the use of variance-mean Gaussian mixtures (VMGM) \cite{ref:Barndorff-Nielsen}. Different approaches have been considered in the literature using VMGM. In \cite{ref:Figueiredo2003}, the $\ell_1$-norm penalty is considered for promoting sparsity, which is interpreted as a Laplace prior for a MAP interpretation of the $\ell_1$ penalty using the EM algorithm. For robust regression, VMGMs can be utilized to express a least $\ell_q$ regression as a ML estimation problem, which is then solved using the EM algorithm \cite{ref:Lange1993}. The convergence properties of the approach in \cite{ref:Lange1993} solved by using iterative re-weighted least squares were presented in \cite{ref:Ba2014}. A general approach for MAP estimation is considered in \cite{ref:Polson2013}, where the main idea for utilizing the VMGMs and EM algorithm for the \textit{prior} pdf was extended to different VMGMs. This idea was further extended in \cite{ref:Godoy2014} for sparse parameter vectors with quantized data, which in turn is a generalization of the work in \cite{ref:Carvajal2012}. In this paper, we provide all the relevant theory for the formulation of the EM algorithm, when VMGMs are employed, for \textit{hidden variables} in both the \textit{likelihood function} and the \textit{prior} pdf tailored to the $\ell_q$-norm, $0<q \leq 1$. To illustrate the effectiveness of our proposal, we present an example with both \textit{hidden variables} and nonlinearities and compare the results with utilizing the $\ell_1$- and the $\ell_q$-norms with both the CD and the EM-based MAP approaches.

The remainder of the paper is organized as follows: Section \ref{sec:EM-basic} introduces the basic EM definitions for the ML, penalized ML and MAP estimation problems. The derivation of the ECM algorithm with the $\ell_q$-norm penalty is included. Section \ref{sec:Mixtures} introduces the basic notion of mixtures for expressing a prior distribution and provides all the necessary expressions for the formulation of the EM algorithm, when the prior is expressed as a VMGM. Section \ref{sec:Sparse} presents the VMGM expressions for the $\ell_q-$norm penalty. Finally, simulation results are provided in Section \ref{sec:sims}, while Section \ref{sec:concl} concludes the paper.


\section{Basic Description of the EM algorithm}
\label{sec:EM-basic}

The EM algorithm is basically formulated to iteratively solve the ML estimation problem:
\begin{align}\label{eq:MLprob}
\bs{\hat{\theta}}_{\rm ML}=\arg\max_{\bs{\theta}} \mathcal{L}(\bs{\theta}).
\end{align}
Using the notion of a \textit{hidden} variable $\bs{z}$ \cite{ref:McLachlan}, we can easily express the log-likelihood function as follows:
\begin{equation}\label{eq:loglik1}
\mathcal{L}(\bs{\theta})=\mathcal{Q}_{\rm ML}\left(\bs{\theta},\bs{\hat{\theta}}^{(i)}\right)-\mathcal{H}_{\rm ML}\left(\bs{\theta},\bs{\hat{\theta}}^{(i)}\right),
\end{equation}
where $\bs{\hat{\theta}}^{(i)}$ is the estimate of $\bs \theta$ at the $i$th iteration, and 
\begin{align*}
\mathcal{Q}_{\rm ML}\left(\bs{\theta},\bs{\hat{\theta}}^{(i)}\right)&=\int \log p(\bs{z},\bs{y}|\bs{\theta})p\left(\bs{z}|\bs{y},\bs{\hat{\theta}}^{(i)}\right)d\bs{z},\\
\mathcal{H}_{\rm ML}\left(\bs{\theta},\bs{\hat{\theta}}^{(i)}\right)&=\int \log p(\bs{z}|\bs{y},\bs{\theta})p\left(\bs{z}|\bs{y},\bs{\hat{\theta}}^{(i)}\right)d\bs{z}.
\end{align*}
Using Jensen's inequality \cite{ref:Bishop}, one can readily see that \cite{ref:Dempster1977} \[\mathcal{H}_{\rm ML}\left(\bs{\theta},\bs{\hat{\theta}}^{(i)}\right) \leq \mathcal{H}_{\rm ML}\left(\bs{\hat{\theta}}^{(i)},\bs{\hat{\theta}}^{(i)}\right)\]for any $\bs{\theta}$. Therefore, the EM algorithm for the ML estimation problem can be formulated as follows \cite{ref:McLachlan}:\vspace{1mm}

\noindent \textbf{E-step}: Compute\vspace{-2mm}
 \[
 \mathcal{Q}_{\rm ML}\left(\bs{\theta},\bs{\hat{\theta}}^{(i)}\right)=E\left[\log p(\bs{z},\bs{y}|\bs{\theta})|\bs{y},\bs{\hat{\theta}}^{(i)}\right].\vspace{-2mm}
 \]

\noindent \textbf{M-step}: Solve\vspace{-2mm}
 \[
 \bs{\hat{\theta}}^{(i+1)}=\arg\max_{\bs{\theta}} \mathcal{Q}_{\rm ML}\left(\bs{\theta},\bs{\hat{\theta}}^{(i)}\right).\vspace{-2mm}
 \]

Under some general conditions, the above algorithm converges to the ML estimate of $\bs{\theta}$ \cite{ref:McLachlan}.

\subsection{Penalized ML estimation via the EM algorithm} \label{sec:EM_pen}

Many regularized estimation problems involve the optimization of a cost function of the form
\begin{equation}
U(\bs{\theta})=\mathcal{L}(\bs{\theta}) - g(\bs{\theta}),
\end{equation}
where $g(\cdot)$ is a penalty function. If we consider that $\mathcal{L}(\bs{\theta})$ is the \textit{log-likelihood} function, then this is a penalized ML estimation problem. For the attainment of the estimates, different approaches can be used, depending on the nature of the penalty function. In our case, in order to promote sparsity on $\bs \theta$, a popular choice for the penalty function is $g(\bs{\theta})=\frac{1}{\tau}\Vert \bs \theta \Vert_q^q$ with $0<q\leq1$. When $q=1$, we obtain the  Lasso \cite{ref:Tibshirani1996}. 
In order to solve the penalized ML estimation problem, the modifications to the EM algorithm are implemented on the M-step, while the E-step remains the same as before \cite[Chap. 5.18]{ref:McLachlan}. Hence, the modified EM algorithm for penalized ML estimation is given as follows:\\
\textbf{E-step}: Compute\vspace{-2mm}
 \[
 \mathcal{Q}_{\rm ML}\left(\bs{\theta},\bs{\hat{\theta}}^{(i)}\right)=E\left[\log p(\bs{z},\bs{y}|\bs{\theta})|\bs{y},\bs{\hat{\theta}}^{(i)}\right].\vspace{-2mm}
 \]

\noindent \textbf{M-step}: Solve\vspace{-2mm}
 \[
 \bs{\hat{\theta}}^{(i+1)}=\arg\max_{\bs{\theta}} \mathcal{Q}_{\rm ML}\left(\bs{\theta},\bs{\hat{\theta}}^{(i)}\right) - g(\bs{\theta}).\vspace{-2mm}
 \]
In order to understand why traditional methods cannot be applied straightforwardly, let us consider a linear regression as in \eqref{eq:lGm}, but where the matrix $\bs X$ is a function of a \textit{hidden variable}, e.g., an unknown state in a state-space representation. In such a case, when performing the E-step, we need to compute $\bs E_1 = E[\bs X | \bs y]$ and $\bs E_2 = E[\bs X^T \bs X | \bs y]$. In general, methods that do not consider \textit{hidden variables} rely on the relationship between the matrix $\bs X$ and $\bs X^T \bs X$. However, in our case, $\bs E_2 \neq \bs E_1^T \bs E_1$, and hence traditional methods should be modified accordingly. 

\subsubsection{The ECM algorithm for $\ell_q$-norm penalized ML estimation}
For non-convex penalties, a CD algorithm \cite{ref:Zangwill} (also called Cyclic Descent) was first proposed and analysed in \cite{ref:Breheny2011} and specifically applied to the $\ell_q$-norm penalty as a sparsity inducing norm in \cite{ref:Marjanovic2014}. The basic idea behind \cite{ref:Breheny2011} and \cite{ref:Marjanovic2014} is to optimize the penalized ML problem in a coordinate-wise manner. This implies computing the gradient of the \textit{log-likelihood} function with respect to one component of $\bs \theta$ and solving for this component, while the remaining components are considered constants. This is implemented for each component of the parameter vector $\bs \theta$ in a cyclic manner until convergence. In our analysis, the system model allows for obtaining an auxiliary function $\mathcal{Q}_{\rm ML}\left(\bs{\theta},\bs{\hat{\theta}}^{(i)}\right)$ that is quadratic with respect to $\bs \theta$. To see this, let us first  assume that the regressor matrix $\bs X$ is a function of a \textit{hidden variable} $\bs z$, $\bs X = f(\bs z)$, and that the measurement noise is zero mean Gaussian distributed with covariance matrix $\bs \Sigma_\epsilon$. With these assumptions, we have:\vspace{-2mm}
\begin{align}
\mathcal{Q}_{\rm ML}\left(\bs{\theta},\bs{\hat{\theta}}^{(i)}\right) & = -\frac{1}{2}\left( \bs y^T \bs \Sigma_\epsilon^{-1}\bs y -2\bs y^T \bs \Sigma_\epsilon^{-1} E[\bs X |\bs{y},\bs{\hat{\theta}}^{(i)} ]\bs \theta \right. \nonumber \\
& \quad \left.  + \, \bs \theta^T E[\bs X^T \bs \Sigma_\epsilon^{-1} \bs X|\bs{y},\bs{\hat{\theta}}^{(i)}]\bs \theta \right), \nonumber\vspace{-2mm}
\end{align}
which is clearly a quadratic function of $\bs \theta$. Then, we compute the gradient with respect to only one component of $\bs \theta$. For simplicity, let us define $\bs E_1 = E[\bs X |\bs{y},\bs{\hat{\theta}}^{(i)} ]$ and $\bs E_2 = E[\bs X^T \bs \Sigma_\epsilon^{-1} \bs X|\bs{y},\bs{\hat{\theta}}^{(i)}]$. With these we obtain \vspace{-2mm}
\begin{align}
\frac{\partial}{\partial \theta_j}\mathcal{Q}_{\rm ML}\left(\bs{\theta},\bs{\hat{\theta}}^{(i)}\right) = \left( [\bs E_1]_j^T\bs \Sigma_\epsilon \bs y -  \textbf{q}_j^T\bs E_2 \bs \theta  \right),\vspace{-2mm}
\label{eq:grad_comp}
\end{align}
where $[\bs E_1]_j$ is the $j$th column of $\bs E_1$ and $\textbf{q}_j$ is the $j$th column of the identity matrix. If we solve for $\theta_j$, we obtain\scriptsize{
\begin{equation*}
\theta_j^{(i+1)^{*}} = \frac{[\bs E_1]_j^T\bs \Sigma_\epsilon^{-1} \bs y - [\bs E_2]_j^T [\theta_1^{(i+1)}, ... ,\theta_{j-1}^{(i+1)}, 0 , \theta_{j+1}^{(i)}, ...,\theta_p^{(i)}]^T}{[\bs E_2]_{j,j}}.
\end{equation*}}
\normalsize
Since we are interested in the penalized ML problem, we include the penalty for the $j$th component of $\bs \theta$ and solve for
\[ \theta_j^{(i+1)} = \text{arg}\max_{\theta_j^{(i+1)}} -\frac{1}{2}\left(\theta_j^{(i+1)} - \theta_j^{(i+1)^{*}}\right)^2 - \frac{\left \vert\theta_j^{(i+1)} \right\vert^q}{\tau [\bs E_2]_{j,j}}, \]
using the scalar optimization algorithm described in \cite{ref:Marjanovic2014}.

\subsection{MAP estimation via the EM algorithm}

Let us consider now the case of the MAP estimation. Based on Bayes' theorem, the \textit{log-posterior} pdf of $\bs \theta$ can be expressed as:
\begin{equation}
\log p(\bs{\theta}|\bs{y})= \mathcal{L}(\bs{\theta})+G(\bs{\theta})-\log p(\bs{y}).
\label{eq:logPost}
\end{equation}

Using (\ref{eq:loglik1}), (\ref{eq:logPost}) and the fact that $\mathcal{H}_{\rm ML}\left(\bs{\theta},\bs{\hat{\theta}}^{(i)}\right)$ is a decreasing function of $\bs \theta$, the EM algorithm for the MAP estimation problem is formulated as follows:

\textbf{E-step}: Compute
 \[
 \mathcal{Q}_{\rm MAP}\left(\bs{\theta},\bs{\hat{\theta}}^{(i)}\right)=E\left[p(\bs{z},\bs{y}|\bs{\theta})|\bs{y},\bs{\hat{\theta}}^{(i)}\right]+G(\bs{\theta}).
 \]

 \textbf{M-step}: Solve
 \[
 \bs{\hat{\theta}}^{(i+1)}=\arg\max_{\bs{\theta}} \mathcal{Q}_{\rm MAP}\left(\bs{\theta},\bs{\hat{\theta}}^{(i)}\right).
 \]

 In the next section, we will see how we can express the \textit{log-prior} so that the EM algorithm can be used to tackle the $\ell_q-$norm regularized ML estimation problem.

\section{MAP: The prior pdf as a mixture of distributions}
\label{sec:Mixtures}


Focusing on the prior distribution for MAP estimation, it is possible to express $p(\bs{\theta})$ as a marginal probability density function (pdf), occurring by integrating out the rest of the
variables in a possible joint pdf. In this sense, one can express $p(\bs{\theta})$ as an infinite mixture, where there is an underlying process that generates the desired pdf for the parameters. More precisely:
\begin{equation}
p(\bs{\theta})=\int p(\bs{\theta}|\bs{\lambda})p(\bs{\lambda})d\bs{\lambda},
\label{eq:VMGM}
\end{equation}
where $\bs{\lambda} \in \mathbb{R}^r$ is interpreted as a \emph{hidden} (hyper)-parameter vector. If the conditional pdf $p(\bs{\theta}|\bs{\lambda})$ is Gaussian, then we obtain a Gaussian mixture, see, e.g., \cite{ref:Barndorff-Nielsen, ref:West1987, ref:Keilson1974}. Hence, we can express the \textit{log-prior} as
\begin{equation}\label{eq:logPrior}
\log p(\bs{\theta})=\mathcal{Q}_{\rm prior}\left(\bs{\theta},\bs{\hat{\theta}}^{(i)}\right)-\mathcal{H}_{\rm prior}\left(\bs{\theta},\bs{\hat{\theta}}^{(i)}\right)
\end{equation}
with
\begin{align*}
\mathcal{Q}_{\rm prior}\left(\bs{\theta},\bs{\hat{\theta}}^{(i)}\right)&=\int \log p(\bs{\theta},\bs{\lambda})p\left(\bs{\lambda}|\bs{\hat{\theta}}^{(i)}\right)d\bs{\lambda},\\
\mathcal{H}_{\rm prior}\left(\bs{\theta},\bs{\hat{\theta}}^{(i)}\right)&=\int \log p(\bs{\lambda}|\bs{\theta})p\left(\bs{\lambda}|\bs{\hat{\theta}}^{(i)}\right)d\bs{\lambda}.
\end{align*}
The auxiliary function $\mathcal{H}_{\rm prior}\left(\bs{\theta},\bs{\hat{\theta}}^{(i)}\right)$ is  decreasing in $\bs{\theta}$\footnote{In the same way as $\mathcal{H}\left(\bs{\theta},\bs{\hat{\theta}}^{(i)}\right)$.}. Therefore, the EM algorithm can be formulated as follows:

\noindent \textbf{E-step}: Compute
 \[
 \mathcal{Q}_{\rm MAP-EM}\left(\bs{\theta},\bs{\hat{\theta}}^{(i)}\right)=\mathcal{Q}_{\rm ML}\left(\bs{\theta},\bs{\hat{\theta}}^{(i)}\right)+\mathcal{Q}_{\rm prior}\left(\bs{\theta},\bs{\hat{\theta}}^{(i)}\right).
 \]
\noindent \textbf{M-step}: Solve
 \[
 \bs{\hat{\theta}}^{(i+1)}=\arg\max_{\bs{\theta}} \mathcal{Q}_{\rm MAP-EM}\left(\bs{\theta},\bs{\hat{\theta}}^{(i)}\right).
 \]
For the M-step of the last EM algorithm, we need to evaluate the gradient of $\mathcal{Q}_{\rm MAP-EM}\left(\bs{\theta},\bs{\hat{\theta}}^{(i)}\right)$ with respect to $\bs{\theta}$. Assuming that $\nabla_{\bs{\theta}}\mathcal{Q}_{\rm ML}\left(\bs{\theta},\bs{\hat{\theta}}^{(i)}\right)$ is known from the ML estimation problem, we only need to evaluate $\nabla_{\bs{\theta}}\mathcal{Q}_{\rm prior}\left(\bs{\theta},\bs{\hat{\theta}}^{(i)}\right)$. 
\begin{lemma}
When expressing the prior pdf of $\bs \theta$, $p(\bs \theta)$, as a VMGM and letting $\bs{\theta}|\bs{\lambda}\sim\mathcal{N}(\bs{\mu}_{\bs{\theta}}(\bs{\lambda}),\bs{\Sigma}_{\bs{\theta}}(\bs{\lambda}))$, the gradient of the auxiliary function $\mathcal{Q}_{\rm prior}\left(\bs{\theta},\bs{\hat{\theta}}^{(i)}\right)$, is given by
\begin{eqnarray}\label{eq:Ggrad}
 \nabla_{\bs{\theta}}G(\bs{\theta})\left.\right|_{\bs{\hat{\theta}}^{(i)}}=&& E_{\bs{\lambda}|\bs{\hat{\theta}}^{(i)}}\left[-\bs{\Sigma}^{-1}_{\bs{\theta}}(\bs{\lambda})\right]\bs{\hat{\theta}}^{(i)}\nonumber\\&&+E_{\bs{\lambda}|\bs{\hat{\theta}}^{(i)}}\left[\bs{\Sigma}^{-1}_{\bs{\theta}}(\bs{\lambda})\bs{\mu}_{\bs{\theta}}(\bs{\lambda})\right].
 \end{eqnarray}
\end{lemma}
\proof
First, notice that the marginal of $\bs{\lambda}$ is independent of $\bs{\theta}$. Therefore,
\[
\nabla_{\bs{\theta}}\mathcal{Q}_{\rm prior}\left(\bs{\theta},\bs{\hat{\theta}}^{(i)}\right)=\int \nabla_{\bs{\theta}}\log p(\bs{\theta}|\bs{\lambda})p\left(\bs{\lambda}|\bs{\hat{\theta}}^{(i)}\right)d\bs{\lambda}.
\]
Given that $\bs{\theta}|\bs{\lambda}\sim \mathcal{N}(\bs{\mu}(\bs{\lambda}),\bs{\Sigma}_{\bs{\theta}}(\bs{\lambda}))$, we obtain:\small{
\begin{align}
\nabla_{\bs{\theta}}\mathcal{Q}_{\rm prior} \left(\bs{\theta},\bs{\hat{\theta}}^{(i)}\right)&= \int \left[-\bs{\Sigma}^{-1}_{\bs{\theta}}(\bs{\lambda})(\bs{\theta}-\bs{\mu}(\bs{\lambda}))\right]p\left(\bs{\lambda}|\bs{\hat{\theta}}^{(i)}\right)d\bs{\lambda} \nonumber \\
&= E_{\bs{\lambda}|\bs{\hat{\theta}}^{(i)}}\left[- \bs{\Sigma}^{-1}_{\bs{\theta}}(\bs{\lambda})\right]\bs{\theta} \nonumber \\& +E_{\bs{\lambda}|\bs{\hat{\theta}}^{(i)}}\left[\bs{\Sigma}^{-1}_{\bs{\theta}}(\bs{\lambda})\bs{\mu}(\bs{\lambda}))\right].
\end{align}}\normalsize
The expected values in the last equation can be computed based on the Gaussianity assumption for $\bs{\theta}|\bs{\lambda}$. Due to this assumption, $p(\bs{\theta})$ satisfies the following equation:
\begin{equation}\label{eq:nablaptheta}
\nabla_{\bs{\theta}}p(\bs{\theta})=\int \left[-\bs{\Sigma}^{-1}_{\bs{\theta}}(\bs{\lambda})(\bs{\theta}-\bs{\mu}(\bs{\lambda}))\right]p\left(\bs{\theta}|\bs{\lambda}\right)p(\bs{\lambda})d\bs{\lambda}.
\end{equation}
On the other hand, Bayes rule yields:
\[
p(\bs{\lambda})=\frac{p(\bs{\lambda}|\bs{\theta})p(\bs{\theta})}{p(\bs{\theta}|\bs{\lambda})}.
\]
Plugging in the last expression into (\ref{eq:nablaptheta}), we obtain:
\[
\nabla_{\bs{\theta}}p(\bs{\theta})=\int \left[-\bs{\Sigma}^{-1}_{\bs{\theta}}(\bs{\lambda})(\bs{\theta}-\bs{\mu}(\bs{\lambda}))\right]p\left(\bs{\lambda}|\bs{\theta}\right)p(\bs{\theta})d\bs{\lambda}.
\]
Therefore,
\begin{equation}\label{eq:Gderivative}
\frac{1}{p(\bs{\theta})}\nabla_{\bs{\theta}}p(\bs{\theta})=E_{\bs{\lambda}|\bs{\theta}}\left[-\bs{\Sigma}^{-1}_{\bs{\theta}}(\bs{\lambda})(\bs{\theta}-\bs{\mu}(\bs{\lambda}))\right].
\end{equation}
We now note that (\ref{eq:Gderivative}) equals $\dot{G}(\bs{\theta})=\nabla_{\bs{\theta}}\log p(\bs{\theta})$. Nevertheless, $G(\bs{\theta})$ is known and so is $\dot{G}(\bs{\theta})$. Therefore, (\ref{eq:Ggrad}) follows. \eop

The expression in \eqref{eq:Ggrad} corresponds to a linear system of equations that can be generally solved, because $p\left(\bs{\lambda}|\bs{\hat{\theta}}^{(i)}\right)$ is a known distribution. Thus, the expected values can be computed either in closed form when possible or using Monte Carlo techniques, such as the Metropolis-Hastings algorithm or the Gibbs sampler \cite{ref:Robert}. A particular case arises when $G(\bs{\theta})$ can be expressed as a function of the coordinate variables of $\bs{\theta}$ as follows:
 \begin{equation}\label{eq:Gsum}
 G(\bs{\theta})=\sum_{j=1}^p \kappa\left(\frac{\theta_j}{\tau}\right).
 \end{equation}
Here, $\tau$ is a factor controlling the strength of the regularization. Different selections of the kernel function $\kappa(\cdot)$ lead to different regularizations. Usual regularization choices can be obtained based on Gaussian mixtures, e.g., the Ridge regression for $\kappa(\theta_j/\tau)=(\theta_j/\tau)^2$ and the Lasso for $\kappa(\theta_j/\tau)=|\theta_j/\tau|$ \cite{ref:Godoy2014}. The general form of a VMGM for a scalar parameter $\theta_j$ is given by\vspace{-2mm}
\begin{equation}\label{eq:VMGMscalar}
p(\theta_j)=\int_{0}^{\infty} \mathcal{N}\left(\mu_j+\lambda u_j, \tau^2\lambda\right)p(\lambda)d\lambda.\vspace{-2mm}
\end{equation}
Based on this formula, the Ridge regression is obtained for $u_j=\mu_j=0$ and $\lambda=1$ and the Lasso is obtained for $u_j=\mu_j=0$ and $\lambda$ exponentially distributed. Clearly, these two very usual regularizations are exactly expressed as VMGMs. For the cases where $\mu_j=0$, the following lemma holds:

\begin{lemma}\label{lem:1}
When $\mu_j=0, j=1,2,\ldots, p$, then\vspace{-2mm}
\begin{equation}
\frac{\partial \mathcal{Q}_{\rm prior}\left(\bs{\theta},\bs{\hat{\theta}}^{(i)}\right)}{\partial\theta_j}=\frac{1}{\theta_j^{(i)}}\kappa(\theta_j)\left.\right|_{\theta_j=\theta_j^{(i)}}\theta_j.\vspace{-2mm}
\end{equation}
\end{lemma}
\proof
See \cite{ref:Godoy2014}. \eop

In the next section, we present in detail the VMGM descriptions of the $\ell_q-$norm regularization term.

\section{Sparse Parameter Estimation using VMGM\MakeLowercase{s}}
\label{sec:Sparse}

As it has already become apparent, the penalties that are usually considered for promoting sparsity can be interpreted as MAP problems. In this section, we explore the representation of the $\ell_q-$norm as a \textit{prior} distribution function for developing a sparse-promoting algorithm under the utilization of VMGMs. To this end, the density function that relates to the $\ell_q-$norm is given by\vspace{-2mm}
\begin{align}
p_{\ell_q}(\bs \theta) = & \prod_{j = 1}^{p}p_{\ell_q}(\theta_j), \label{eq:lq_pdf}\\
p_{\ell_q}(\theta_j) = & \frac{1}{2\Gamma\left( 1+\frac{1}{q} \right)\sqrt{\tau^2}}e^{-\frac{1}{\tau^q} |\theta_j|^q},\vspace{-3mm}
\label{eq:lqj_pdf}
\end{align}
which corresponds to a \textit{leptokurtic} exponential power distribution \cite{ref:Solaro2004}. The prior $p_{\ell_q}(\theta_j)$ in \eqref{eq:lqj_pdf} can be expressed as a VMGM, with $p(\lambda_j)$ being a stable distribution in \eqref{eq:VMGMscalar}. Then, the MAP-EM algorithm can be used to solve the aforementioned sparsity-promoting estimation problems. 

\subsection{MAP-EM algorithm for sparse parameter estimation}
\label{section:MAP-EM}

In this section, we present the relevant expressions in the development of the MAP-EM algorithm for sparse-parameter estimation using the $\ell_q-$norm. 
\begin{lemma}
\label{lemma:dQ_prior_d_thetaj}
In the case that the \textit{log-prior} function is given by \eqref{eq:Gsum} and $p(\theta_j)$ is a VMGM as in \eqref{eq:VMGMscalar}, $\frac{\partial \m{Q}_{\text{prior}}(\bs \theta,\hat{\theta}^{(i)})}{\partial \theta_j},\, \forall j = 1,2,...,p,$ is then given by\vspace{-2mm}
\begin{align}
  \frac{\partial \m{Q}_{\text{prior}}(\bs \theta,\hat{\bs \theta}^{(i)})}{\partial \theta_j}  & = \int \left ( -\frac{\theta_j }{\tau^2 \lambda_j} \right )p(\lambda_j|\hat{\theta}_j^{(i)})d(\lambda_j) \nonumber \\
  & = \left[ - \frac{\theta_j }{\tau^2} {E}[\lambda_j^{-1}|\hat{\theta}_j^{(i)}] \right],\vspace{-4mm}
\label{eq:dqprior_dthetaj}
\end{align}
where ${E}[\lambda_j^{-1}|\hat{\theta}_j^{(i)}]$ is the
expectation obtained from \vspace{-2mm}
\begin{equation}\label{eq:dlog_pen_eq}
- \frac{\hat{\theta}^{(i)}_j }{\tau^2 }{E}[\lambda_j^{-1}|\hat{\theta}_j^{(i)}]  = -\dot{\kappa}\left(\frac{\hat{\theta}^{(i)}_j}{\tau }\right).\vspace{-2mm}
\end{equation}
Equivalently, we obtain\vspace{-2mm}
\begin{equation}
\frac{d \mathcal{Q}_{\rm prior}\left(\bs \theta,\hat{\bs \theta}^{(i)}\right)}{d \bs \theta} = -\bs{K}^{(i)}{\bs \theta},\vspace{-2mm}
\label{eq:dQ_prior_d_theta}
\end{equation}
where \vspace{-2mm}
\begin{equation}
\bs{K}^{(i)} = \text{diag}\left(\frac{1}{\hat{{\theta}}_1^{(i)}}\dot{\kappa}\left(\frac{\hat{{\theta}}_1^{(i)}}{\tau}\right) , \ldots , \frac{1}{\hat{{\theta}}_{p}^{(i)}}\dot{\kappa}\left(\frac{\hat{{\theta}}_{p}^{(i)}}{\tau}\right)   \right).\vspace{-2mm}
\label{eq:Kappa}
\end{equation}
\end{lemma}
\proof See \cite{ref:Godoy2014}.
\eop
Since the system is defined as a linear regression, it yields the derivative of $\m{Q}_\text{ML}(\bs{\theta},\hat{\bs{\theta}}^{(i)})$ with respect to $\bs \theta$ in the form:\vspace{-1mm}
\begin{equation}\small{
\frac{d \mathcal{Q}_{\text{ML}}(\bs{\theta},\hat{\bs{\theta}}^{(i)})}{d \bs \theta}  = \frac{{E[\bs{X}^T \vert  \bs{y}, \hat{\bs{\theta}}^{(i)}]\bs{y} - E[ \bs{X}^T\bs{X} \vert \bs{y}, \hat{\bs{\theta}}^{(i)} ]\bs \theta}}{\sigma^2}.}\vspace{-1mm}
\label{eq:dQ_ML_d_theta}
\end{equation}
Then, combining \eqref{eq:dqprior_dthetaj} and \eqref{eq:dQ_ML_d_theta}, we obtain the derivative of $\m{Q}_\text{MAP-EM}(\bs{\theta},\hat{\bs{\theta}}^{(i)})$ with respect to $\bs \theta$ as:\vspace{-2mm}
\begin{equation}
\frac{d \mathcal{Q}_{\text{MAP-EM}}(\bs{\theta},\hat{\bs{\theta}}^{(i)})}{d \bs \theta} = - (\bs B^{(i)} + \bs K^{(i)})\bs \theta + \bs{a}^{(i)} ,\vspace{-2mm}
\label{eq:dQ_MAP-EM_d_theta}
\end{equation}
where $\bs B^{(i)} \!= {E[ \bs{X}^T\bs{X} \vert \bs y, \hat{\bs{\theta}}^{(i)} ]}/\sigma^2$, $\bs K^{(i)}$ is defined in \eqref{eq:Kappa} and $\bs a^{(i)} \!=\! {E[\bs{X}^T \vert  \bs y, \hat{\bs{\theta}}^{(i)}]\bs y}/\sigma^2$.
In particular, the (component-wise) derivative in \textit{Lemma} \ref{lemma:dQ_prior_d_thetaj} for the $\ell_q-$norm of $\bs \theta$ is given by\vspace{-2mm}
\begin{equation}
  \frac{\partial \m{Q}_{\text{prior}}(\bs \theta,\hat{\bs \theta}^{(i)})}{\partial \theta_j} = -\frac{q}{\tau}\left\vert \frac{\hat{\theta}_j^{(i)}}{\tau} \right \vert^{q-1} \frac{\text{sign}(\hat{\theta}_j^{(i)})}{\hat{\theta}_j^{(i)}}\theta_j.\vspace{-2mm}
\label{eq:dqprior_dhj_q}
\end{equation}
\begin{remark}
Notice that the MAP-EM algorithm yields an optimization problem whose solution is given by a system of linear equations, and hence, it is easy to solve\footnote{When compared to the equivalent nonlinear problems obtained by directly applying Lasso or the bridge regression \cite{ref:Fan2010} with $0 < q < 1$.}. This is due to the quadratic form of $\m{Q}_{\text{prior}}(\bs \theta, \hat{\bs \theta}^{(i)})$ with respect to $\bs \theta$. 
\eor \vspace{-4mm}
\end{remark}
\begin{remark}[\hspace{-0.3mm}\cite{ref:Godoy2014}]\label{remark:der_abs_val}
  In our approach, we have to deal with derivatives of functions of
  the form $d| a | / d\, a$, and the derivative at zero does not
  exist. However, it is possible to obtain the desired value at almost
  all points using \vspace{-2mm}
\begin{equation}\label{eq:a_0}
 \frac{d |a|}{d\,a} = \text{sign}(a),\quad \forall\; a \neq 0.\vspace{-2mm}
\end{equation}
Most of the algorithms promoting sparsity have to deal with
non-differentiable cost functions, see,
e.g., \cite{ref:Fan2010,ref:Zou2008}, where a local approximation of the
$\ell_1$-norm is obtained. A common solution to overcome the issue of
non-differentiability when $a \rightarrow 0$ in \eqref{eq:a_0}, is to
start the algorithm from a value far from zero, and then remove the
component $a$ from the parameter vector when it enters in a neighbourhood (defined by the user) of $a=0$ \cite{ref:Figueiredo2003,ref:Polson2013}. \eor \vspace{-4mm}
\end{remark}

\section{Simulations}
\label{sec:sims}

In this section, we provide a numerical illustration for the applicability of our framework. For the simulations, we consider a limited number of measurement points, since,  in general, MAP estimates converge to ML estimates as the number of data points increases. This is known as the \emph{Bernstein-Von Mises theorem} \cite{ref:Lehmann}. In our experience, we have been able to observe this situation for the MAP-EM algorithm with the $\ell_q$-norm, $0 < q \leq 1$.

For the example, we consider the following system model:\vspace{-3mm}
\begin{align*}
z_{k+1} = & \,0.9z_k + \xi_k,  \\
y_k = &\sin \left( \frac{\omega (k-1)}{N}+z_k\right) \bs{u}_{k:k+p-1}^T \bs \theta + \eta_k, \vspace{-5mm}
\end{align*}
where $\bs{u}_{k:k+p-1} = [u_k, u_{k+1},...,u_{k+p-1}]^T$ is a known input, $\theta = [-0.77,-1.55,0,0,0,0,0.46]^T$ is the unknown sparse parameter vector, $N = 256$ is the number of measurement points, $\omega = 5$ is a known constant parameter,  $\xi_k \sim \m{N}(0,0.1)$, and $\eta_k \sim \m{N}(0,0.1)$. For the estimation of $\bs \theta$, we consider the $\ell_q$-norm as presented in this paper, with $q = 0.1$ and $\tau = 0.1$. For comparison purposes, we also consider the corresponding penalized ML problem and solve for it using the EM framework in Section \ref{sec:EM_pen} with $q = 0.1$ and $\tau = 0.1$ as well. Finally, we also consider the approach in \cite{ref:Godoy2014} with $\tau=1$, which corresponds to $q = 1$. Notice that the initial estimate was the same for the four estimation techniques.

\begin{figure}[t]
	\centering
	\includegraphics[width=0.75\textwidth, trim=45mm 2mm 32mm 8mm, clip]{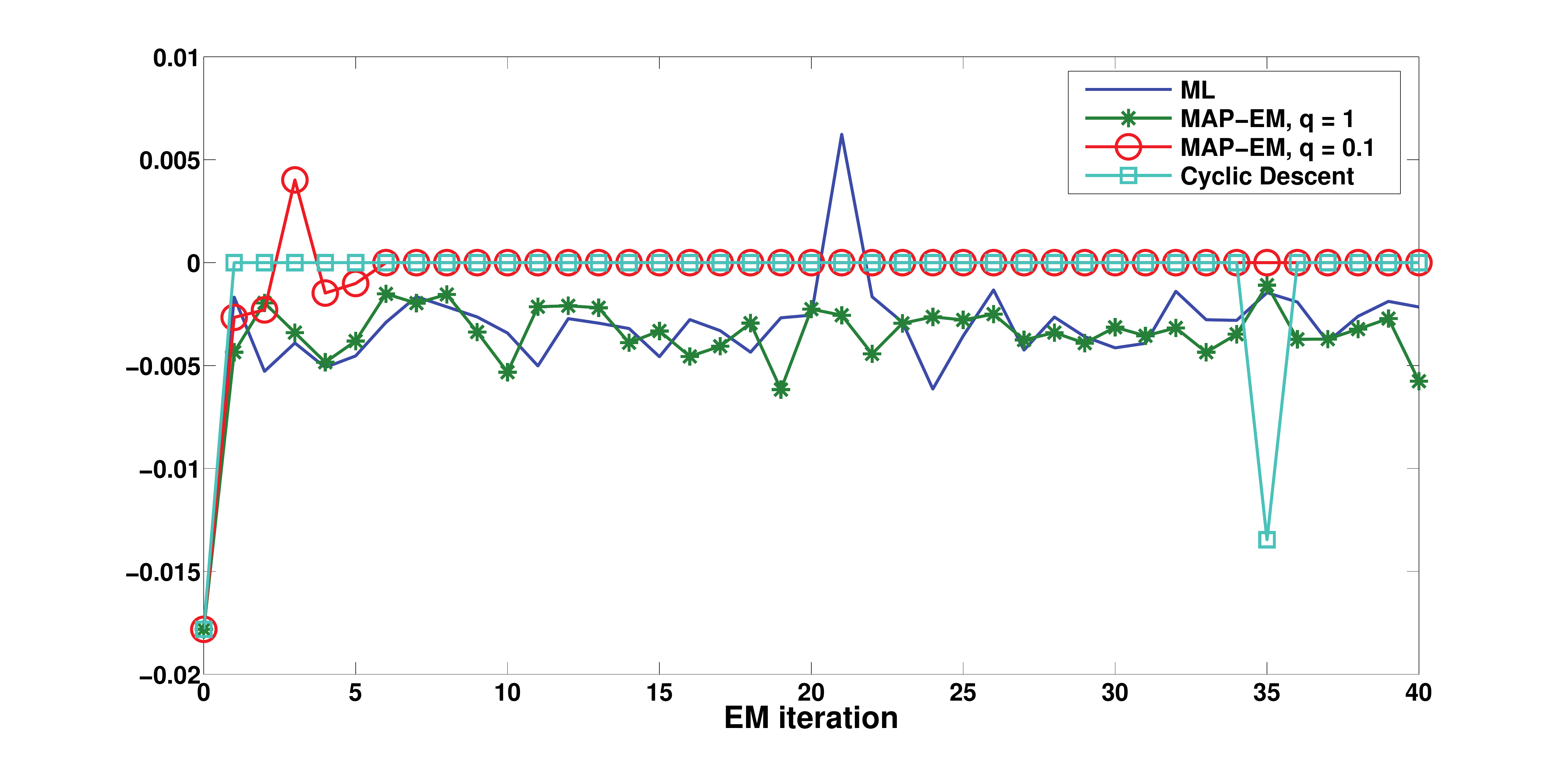}
    \vspace{-6mm}
	\caption{Convergence of the estimates of the first zero of $\bs \theta$ per iteration of the EM algorithm.}
\label{fig:z1}
\end{figure} 
\begin{figure}[t]
	\centering
	\includegraphics[width=0.75\textwidth, trim=45mm 2mm 32mm 8mm, clip]{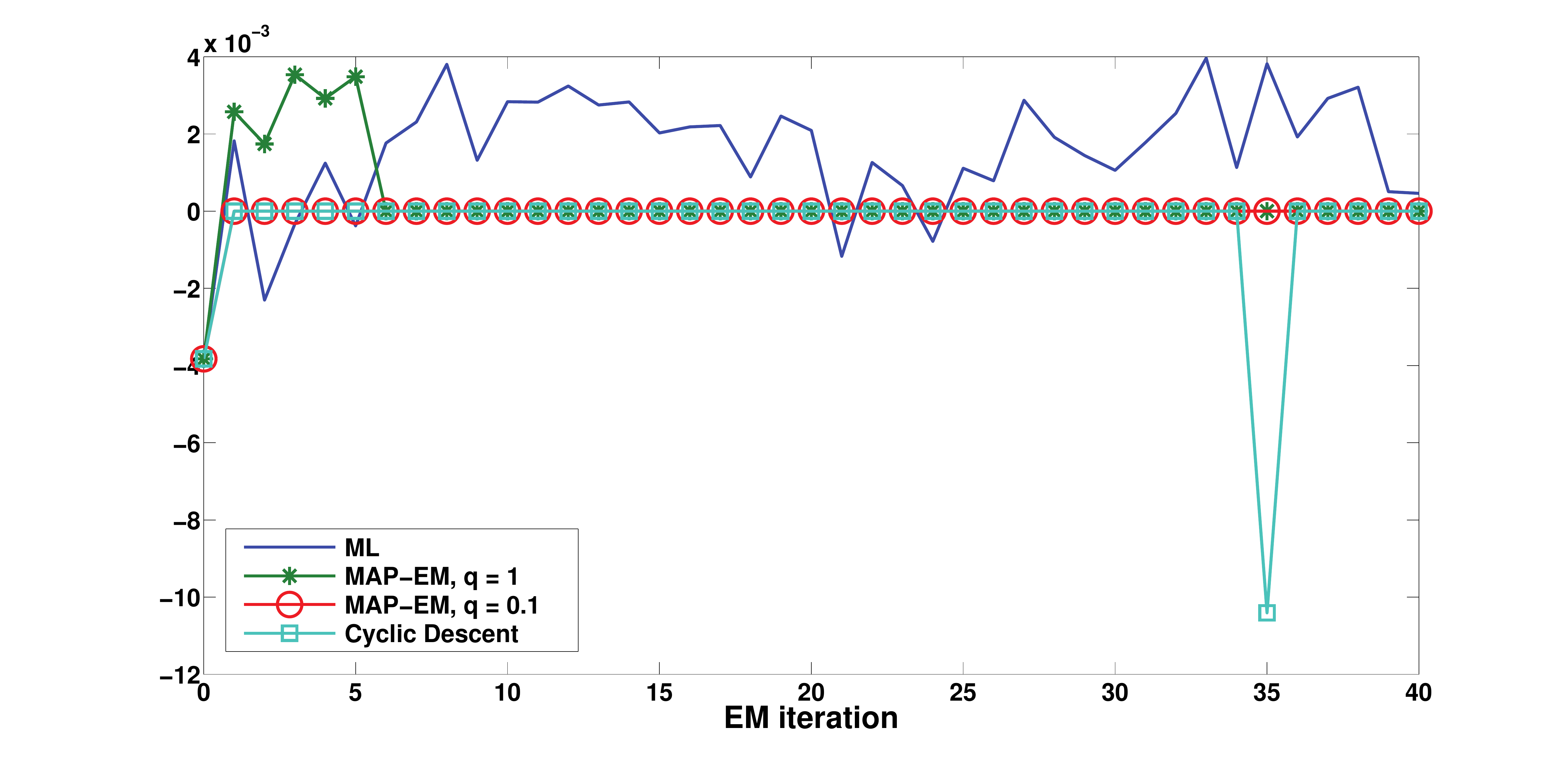}
    \vspace{-6mm}
	\caption{Convergence of the estimates of the second zero of $\bs \theta$ per iteration of the EM algorithm.}
\label{fig:z2}
\end{figure} 
\begin{figure}[t]
	\centering
	\includegraphics[width=0.75\textwidth, trim=45mm 2mm 32mm 8mm, clip]{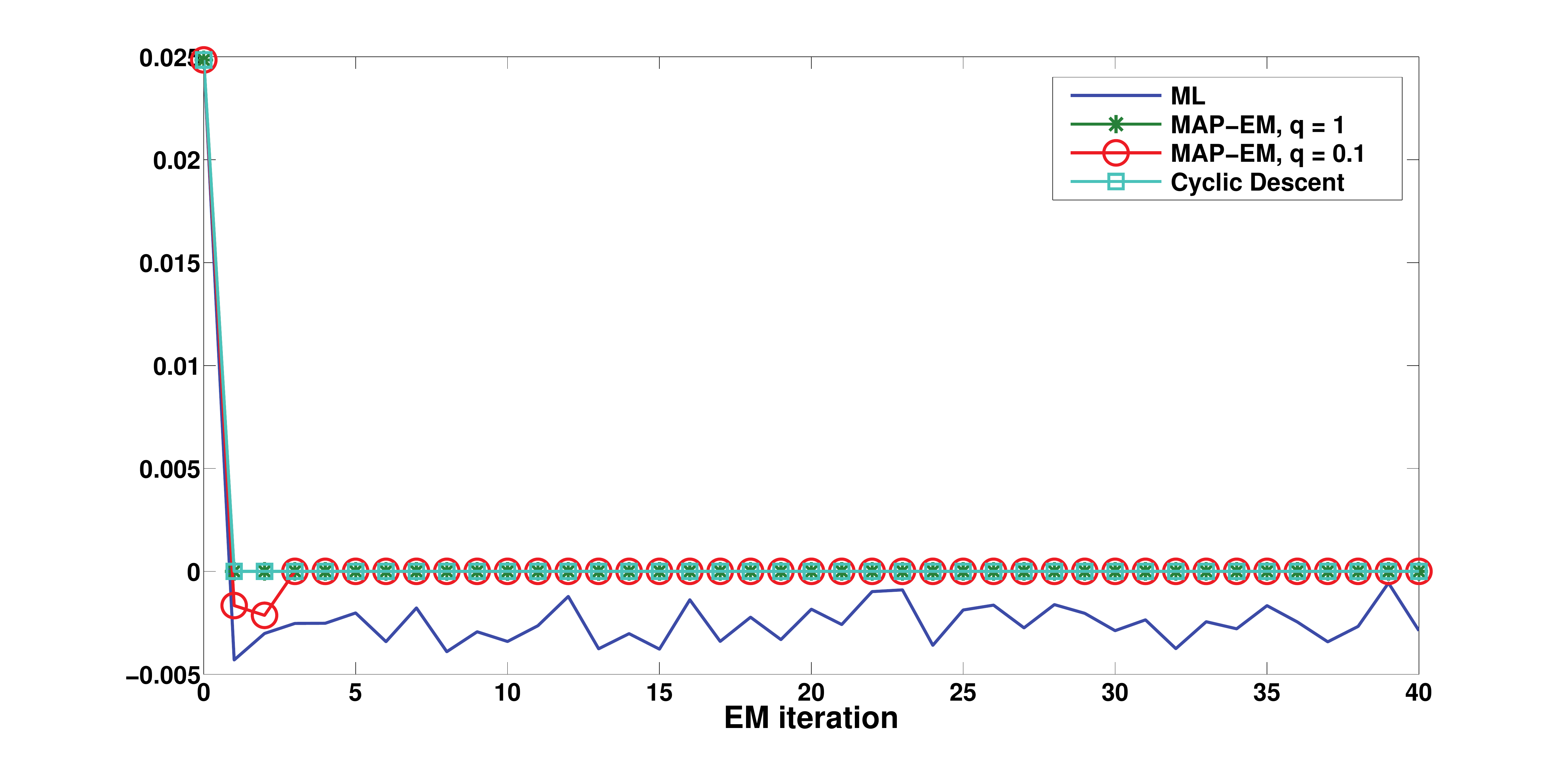}
    \vspace{-6mm}
	\caption{Convergence of the estimates of the third zero of $\bs \theta$ per iteration of the EM algorithm.}
\label{fig:z3}
\end{figure} 
\begin{figure}[t]
	\centering
	\includegraphics[width=0.75\textwidth, trim=45mm 2mm 32mm 8mm, clip]{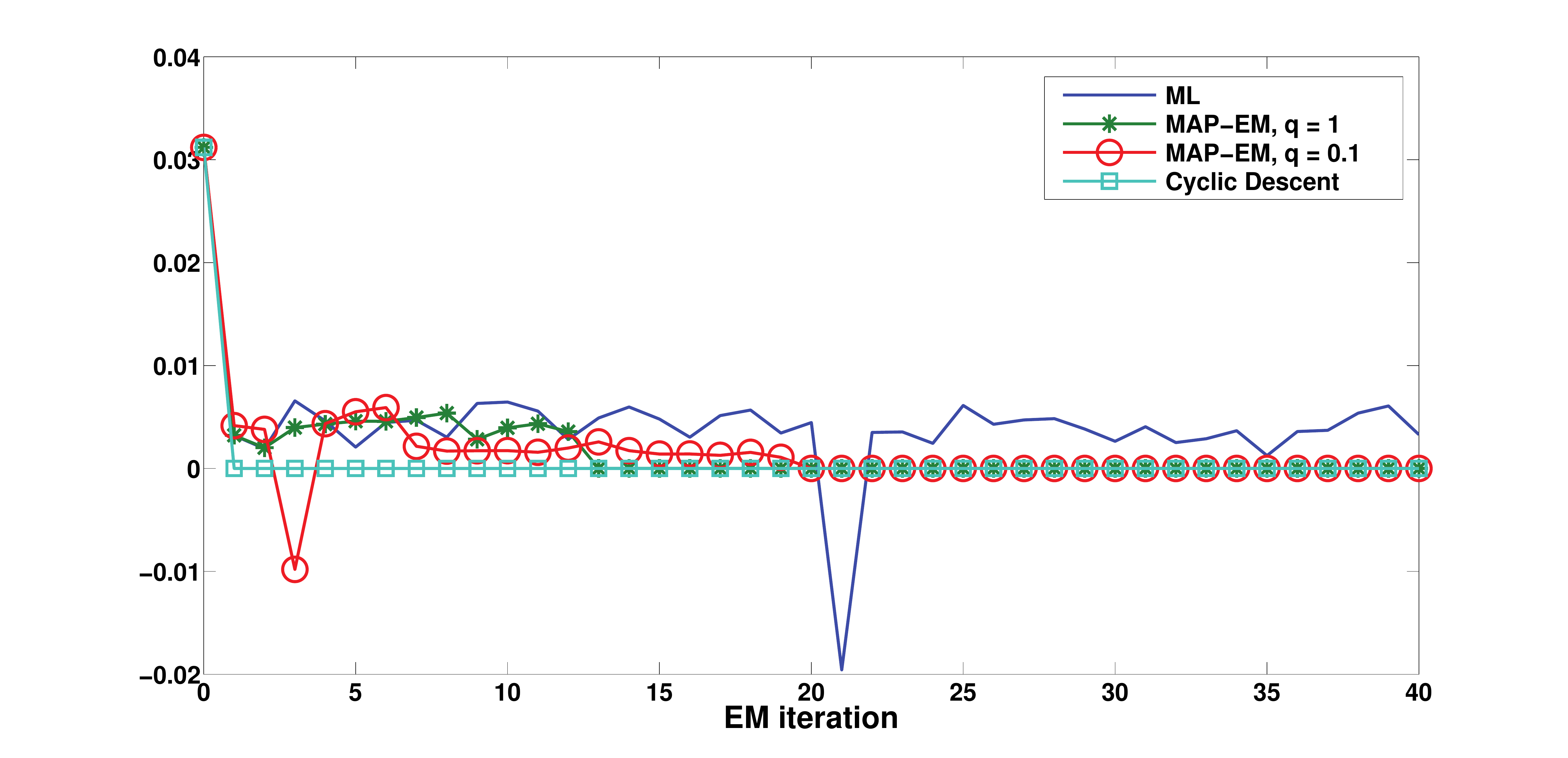}
    \vspace{-6mm}
	\caption{Convergence of the estimates of the fourth zero of $\bs \theta$ per iteration of the EM algorithm.}
\label{fig:z4}
\end{figure} 
The results of the estimation algorithms are shown in Figs. \ref{fig:z1}---\ref{fig:z4}. From those figures, we can clearly see that the proposed approach can effectively estimate all the zeros of $\bs \theta$. However, the convergence is not achieved at the same time for each zero-entry of $\bs \theta$. Our approach clearly outperforms the $\ell_1$-norm MAP-EM approach and ML. In contrast, the CD approach can accurately estimate the zeros of $\bs \theta$ at the first iteration of the EM algorithm. However, this estimation can vary from iteration to iteration, since the optimization procedure is carried out globally for the a single coordinate. As mentioned in \cite{ref:Breheny2011}, this kind of optimization should be carried out locally. From Figs. \ref{fig:z1} and \ref{fig:z2}, we can see that at iteration no. 34, the zeros are correctly estimated. However, at iteration no. 35, those estimates are no longer zero. In contrast, our proposed method forces the zero value once the estimate converges to zero, following the ideas in \textit{Remark} \ref{remark:der_abs_val}. In addition, the mean square error (MSE) of the estimates are: i) $3.56\times10^{-5}$ for ML, ii) $1.124\times10^{-4}$ for MAP-EM with $q=1$, iii) $1.96\times10^{-5}$ for MAP-EM with $q=0.1$, and iv) $3.48\times10^{-5}$ for CD. The difference in MSE is due to the slow convergence of the non-zero elements that CD exhibits. Thus, we believe that our estimation algorithm could be initialized with one iteration of CD for quicker convergence of the EM algorithm.

\section{Conclusions}
\label{sec:concl}
In this paper, we have described a general framework for the implementation of the EM algorithm based on mean-variance Gaussian mixtures, for promoting sparsity via $\ell_q-$norm regularization ($0 < q \leq 1$). Our proposal allows a (computationally-efficient) simple way of handling the sparsity estimation problem, yielding closed form expressions for the EM algorithm. A particular case of the proposed approach includes $q = 1$ (Lasso). The numerical illustration shows the effectiveness of our approach, outperforming both the ML and the Lasso estimators when choosing $q<1$. In addition, the simulations show that our approach can obtain better estimates than Coordinate Descent within the EM framework. Since the Coordinate Descent algorithm is capable of finding the zeros quickly but may diverge, it can be used to initialize our approach for faster convergence.
\vspace{-2mm}
\bibliographystyle{plain}
\bibliography{refs}
\end{document}